\documentclass[a4paper]{jpconf}
\usepackage{graphicx}

\def\NIMA#1#2#3{Nucl. Inst. Methods {\bf A#1} (#2) #3}

\begin{document}
\boldmath
\title{Recent results from the NA48 experiment at CERN:
CP violation and CKM parameter $|V_{us}|$} \unboldmath

\author{Evgueni Goudzovski}

\address{Scuola Normale Superiore, Piazza dei Cavalieri 7, I-56100 Pisa, Italy}

\ead{goudzovs@mail.cern.ch}

\begin{abstract}
Several recent results from the NA48 experiment are presented: a
measurement $|\eta_{+-}|$, search for CP violating phenomena in
$K^\pm\to3\pi$ decays, and a measurement of $|V_{us}|$.
\end{abstract}

\section{Introduction}
The NA48 series of experiments represents the long-term CERN program
in experimental kaon physics. During the 10 years of operation since
1997, several physics programs were accomplished. The experimental
setup has been upgraded in the course of operation; its principal
components are a kaon beam line and a vacuum decay volume followed
by a magnetic spectrometer consisting of four drift chambers, a
trigger scintillator hodoscope, a liquid krypton electromagnetic
calorimeter, and a muon detector~\cite{fa07}.

The present paper contains a number recent precise measurements
based on various data sets: 1) measurement of the indirect CP
violation parameter $|\eta_{+-}|$ with $K_L\to\pi^+\pi^-$ decays; 2)
measurement of the direct CP violating Dalitz plot slope asymmetries
$A_g$ in $K^\pm\to3\pi^\pm$ and $K^\pm\to\pi^\pm\pi^0\pi^0$ decays;
3) measurement of the CKM matrix element $|V_{us}|$ based on partial
widths of the semileptonic $K^\pm\to\pi^0l^\pm\nu$ decays.

\boldmath
\section{Measurement of the indirect CP violation parameter
$|\eta_{+-}|$} \unboldmath

The interest in the measurement of the parameter
$|\eta_{+-}|=A(K_L\to\pi^+\pi^-)/A(K_S\to\pi^+\pi^-)$ stems, in
particular, from the fact that precision measurements of this value
by KTeV and KLOE experiments published in 2004 and 2006,
respectively, were disagreement with the previous world average by
5\%, or more the four standard deviations.

The NA48 measurement if $|\eta_{+-}|$~\cite{la07} is based on the
data taken during 2 days of dedicated running in 1999. The directly
measured quantity is the ratio of the decay rates
$R=\Gamma(K_L\to\pi^+\pi^-)/\Gamma(K_L\to\pi e\nu)$; these decays
are characterized by similar signatures involving two reconstructed
tracks of charged particles. Then $|\eta_{+-}|$ is computed as
\begin{equation}
\label{eta-main} |\eta_{+-}| =
\sqrt{\frac{\Gamma(K_L\to\pi^+\pi^-)}{\Gamma(K_S\to\pi^+\pi^-)}}=
\sqrt{\frac{\textrm{BR}(K_L\to\pi^+\pi^-)}{\textrm{BR}(K_S\to\pi^+\pi^-)}
\cdot\frac{\tau_{KS}}{\tau_{KL}}}.
\end{equation}
In this approach the $K_L$ and $K_S$ lifetimes $\tau_{KL}$ and
$\tau_{KS}$, and the branching fractions $\textrm{BR}(K_L\to\pi
e\nu)$ and $\textrm{BR}(K_S\to\pi^+\pi^-)$ are external inputs taken
from the best single measurements.

The data sample contains about $80\times10^6$ 2-track triggers.
Event selection is rather similar for the $K_L\to\pi^+\pi^-$ and
$K_L\to\pi e\nu$ modes. A crucial difference is electron vs pion
identification based on the ratio of energy deposition in the
electromagnetic calorimeter to track momentum measured by the
spectrometer. Identification efficiency was measured and corrected
for.

Samples of $47\times 10^3$ $K_L\to\pi^+\pi^-$ and $5.0\times 10^6$
$K_L\to\pi e\nu$ candidates were selected, with about $0.5\%$
background contamination in each. Acceptance corrections and
background subtraction were performed by Monte Carlo simulation.
Trigger efficiencies were measured directly with the data and
corrected for. The most relevant systematic uncertainties come from
precision of simulation of kaon energy spectrum, precision of
radiative corrections, and precision of trigger efficiency
measurement. The final result is
\begin{equation}
\Gamma(K_L\to\pi^+\pi^-)/\Gamma(K_L\to\pi e\nu) =
(4.835\pm0.022_{stat.}\pm0.016_{syst.})\times 10^{-3}.
\end{equation}
This leads, subtracting the $K_L\to\pi^+\pi^-\gamma$ direct emission
contribution, but retaining the inner bremsstrahlung contribution,
to
\begin{equation}
\textrm{BR}(K_L\to\pi^+\pi^-) = (1.941\pm0.019)\times 10^{-3}.
\end{equation}
Finally, the CP violating parameter is computed according to
(\ref{eta-main}) to be
\begin{equation}
|\eta_{+-}| = (2.223\pm0.012)\times 10^{-3}.
\end{equation}
The result in in agreement with the recent KLOE and KTeV
measurements, while in contradiction to the world average as of
2004.

\boldmath
\section{Measurement of the direct CP violation parameter $A_g$ in
$K^\pm\to3\pi$ decays}
\unboldmath

The $K^\pm\to\pi^\pm\pi^+\pi^-$ and $K^\pm\to\pi^\pm\pi^0\pi^0$
decays are among the most promising processes in kaon physics for a
search for CP violating phenomena. The $K^\pm\to3\pi$ matrix element
squared is conventionally parameterized by a polynomial expansion
\begin{equation}
|M(u,v)|^2\sim 1+gu+hu^2+kv^2, \label{slopes}
\end{equation}
where $g$, $h$, $k$ are the so called linear and quadratic Dalitz
plot slope parameters ($|h|,|k|\ll |g|$), and the two Lorentz
invariant kinematic variables $u$ and $v$ are defined as
\begin{equation}
u=\frac{s_3-s_0}{m_\pi^2},~~v=\frac{s_2-s_1}{m_\pi^2},~~
s_i=(P_K-P_i)^2,~i=1,2,3;~~s_0=\frac{s_1+s_2+s_3}{3}. \label{uvdef}
\end{equation}
Here $m_\pi$ is the charged pion mass, $P_K$ and $P_i$ are the kaon
and pion four-momenta, the indices $i=1,2$ correspond to the two
pions of the same electrical charge, and the index $i=3$ to the pion
of different charge. A non-zero difference $\Delta g$ between the
slope parameters $g^+$ and $g^-$ describing the decays of $K^+$ and
$K^-$, respectively, is a manifestation of direct CP violation
expressed by the corresponding slope asymmetry
\begin{equation}
A_g = (g^+ - g^-)/(g^+ + g^-) \approx \Delta g/(2g). \label{agdef}
\end{equation}
The above slope asymmetry is expected to be strongly enhanced with
respect to the asymmetry of integrated decay rates. A recent full
next-to-leading order ChPT computation~\cite{ga03} predicts $A_g$ to
be of the order of $10^{-5}$ within the SM. Calculations involving
processes beyond the SM~\cite{sh98,ga00} allow a wider range of
$A_g$, including substantial enhancements up to a few $10^{-4}$.

A measurement of the quantity $A_g$ was performed with a record data
sample collected in 2003--04 with simultaneous $K^+$ and $K^-$
beams~\cite{ba07}. The measurement method exploits cancellations of
major systematic effects due to simultaneous beams and regular
inversions of magnetic fields in the beam line and setup. The
samples of events selected are $3.11\times 10^9$
$K^\pm\to\pi^\pm\pi^+\pi^-$ candidates, and $9.13\times 10^7$
$K^\pm\to\pi^\pm\pi^0\pi^0$ candidates, practically background-free.

The CP violating charge asymmetries of the linear slope parameter of
the Dalitz plot of the $K^\pm\to\pi^\pm\pi^+\pi^-$ and
$K^\pm\to\pi^\pm\pi^0\pi^0$ decays were found to be
\begin{equation}
\begin{array}{rcrllllcrcl}
A_g^c &=& (-1.5 &\pm& 1.5_{stat.} &\pm& 1.6_{syst.})\times
10^{-4} &=& (-1.5&\pm&2.2)\times 10^{-4},\\
A_g^n &=& (1.8 &\pm& 1.7_{stat.}&\pm& 0.6_{syst.})\times 10^{-4}&=&
(1.8&\pm&1.8)\times 10^{-4}.
\end{array}
\end{equation}
The archived precision is more than an order of magnitude better
that those of the previous measurements. The results do not show
evidences for large enhancements due to non-SM physics, and can be
used to constrain extensions of the SM predicting large CP violating
effects.

\boldmath
\section{Measurement of the CKM parameter $|V_{us}|$ with the $K^\pm\to\pi^0l^\pm\nu$ decays}
\unboldmath

Precise measurements of the CKM matrix parameter $|V_{us}|$ are of
interest for tests of CKM unitarity. The $K_{l3}^\pm$ decay rates,
including the internal bremssrahlung process, is given
by~\cite{le84}
\begin{equation}
\Gamma(K_{l3(\gamma)}) = \frac{G_F^2m_K^5}{384\pi^3}S_{EW}|V_{us}|^2
|f_+(0)|^2I_K(1+\delta_K),
\end{equation}
where $f_+(0)$ is a form factor at $q^2=0$, $S_{EW}=1.0232$ is a
short distance electroweak correction, $I_K$ is the phase space
integral depending on form factors, $(1+\delta_K)$ is a
long-distance correction.

The analysis is based on a measurement of the ratios
$\textrm{BR}(K^\pm\to\pi^0e^\pm\nu)/\textrm{BR}(K^\pm\to\pi^\pm\pi^0)$
and
$\textrm{BR}(K^\pm\to\pi^0\mu^\pm\nu)/\textrm{BR}(K^\pm\to\pi^\pm\pi^0)$
using the data collected during 3 days of a dedicated run in
2003~\cite{ba07vus}. The data samples are $87\times10^3$ $K_{e3}$
candidates with 0.02\% background, and $77\times10^3$ $K_{\mu 3}$
candidates with 0.2\% background. The following partial widths are
measured:
\begin{equation}
\begin{array}{rcrcrcrcr}
\mathrm{BR}(K_{e3})&=&0.05168&\pm&0.00019_{stat.}&\pm&0.00008_{syst.}&\pm&0.00030_{norm.},\\
\mathrm{BR}(K_{\mu3})&=&0.03425&\pm&0.00013_{stat.}&\pm&0.00006_{syst.}&\pm&0.00020_{norm.}.
\end{array}
\end{equation}
Here the last (dominating and correlated) uncertainties are due to
precision of the external input $\textrm{BR}(K_{2\pi})$. The
following values of $|V_{us}|f_+(0)$ were computed from $K_{e3}$ and
$K_{\mu 3}$ decays:
\begin{equation}
\begin{array}{rclcll}
|V_{us}|f_+(0)&=&0.2193&\pm&0.0012,&~[K_{e3}]\\
|V_{us}|f_+(0)&=&0.2177&\pm&0.0013.&~[K_{\mu 3}]
\end{array}
\end{equation}
Here the dominating contribution to the uncertainty comes from the
long-distance corrections. Combining these results assuming $\mu-e$
universality and using $f_+(0)=0.961\pm0.008$~\cite{le84}, it is
obtained:
\begin{equation}
|V_{us}|=0.2277\pm0.0013\pm0.0019_{theor.},
\end{equation}
where the second and largest uncertainty owes to the precision of
$f_+(0)$ computation. The above measurement is found to be
consistent with unitarity of the CKM mixing matrix~\cite{ba07vus}.

\section*{Conclusions}
A number of recent kaon measurements by the NA48 collaboration at
CERN were presented. The achieved precisions are similar to or
better than the best previous ones.



\end{document}